\documentclass{eas}
\usepackage{graphicx}

\begin{document}

%%-----------------------------
%%      the top matter
%%-----------------------------

\title{A large-scale survey for variable stars in M~33}
\author{D. Bersier}\address{Astrophysics Research Institute, Liverpool
John Moores University, Birkenhead CH41 1LD, UK}
\author{J.~D.~Hartman}\address{Harvard-Smithsonian Center for
Astrophysics, 60 Garden St., Cambridge, MA~02138, USA}
\author{K.~Z.~Stanek}\address{Department of Astronomy, The Ohio State
University, Columbus, OH~43210, USA}
\author{J.-P.~Beaulieu}\address{Institut d'Astrophysique de Paris,
UMR7095~CNRS, Universit\'e Pierre \& Marie Curie, 98 bis boulevard
Arago, 75014~Paris, France}
\author{J.~Kaluzny}\address{Nicolaus Copernicus Astronomical Center,
ul.~Bartycka 18, 00-716 Warszawa, Poland}
\author{J.-B.~Marquette}\sameaddress{4}
\author{A. Schwarzenberg-Czerny}\sameaddress{5}
\author{V. Scowcroft}\sameaddress{1}
\author{P.~B.~Stetson}\address{Dominion Astrophysical Observatory,
Herzberg Institute of Astrophysics, National Research Council, 5071
West Saanich Road, Victoria, BC V9E 2E7, Canada}

\runningtitle{Variable stars in M~33}

\begin{abstract}

We have started a survey of M~33 in order to find variable stars and
Cepheids in particular. We have obtained more than 30 epochs of
$g^{\prime} r^{\prime} i^{\prime}$ data with the CFHT and the
one-square-degree camera MegaCam.  We present first results from this
survey, including the search for variable objects and a basic
characterization of the various groups of variable stars.

\end{abstract}

\maketitle

\section{Introduction}

The effect of metallicity on the luminosity of Cepheids (and its
subsequent impact on the distance scale) is not well known. A number
of efforts in the last decade have shown that this effect is not very
large (e.g. Sasselov et al. 1997; Kochanek 1997; Kennicutt et
al. 1998; Sakai et al. 2004, Macri et al. 2006). It can, however, have
a significant impact on the value of the Hubble constant when it is
measured via Cepheids.  What has been demonstrated by these recent
works however is what procedure to follow.  First, one needs a large
number of Cepheids. Second, we need to be able to separate the effects
of reddening and metallicity, which tend to have almost an
indistinguishable impact on a Cepheid's energy distribution when
observed through a relatively small wavelength interval; hence a large
wavelength baseline is needed. Third, all Cepheids should be at the
same distance to avoid introducing possible systematic errors.
Fourth, Cepheids in the sample should cover a range of metallicities.
Thus there is a need for a large-scale survey of a nearby galaxy.
This will provide the needed number of Cepheids and fulfill all
conditions required to address this problem.

Incidentally, lots of other things would be made possible by such a
survey.  Microlensing surveys have shown that completely new science
can be done with very large, homogeneous samples of variable stars.
For instance, large numbers of double-mode Cepheids have been found in
the Magellanic Clouds (e.g. \cite{alcock95}). Another example is the
discovery of several Period--Luminosity relations for red giants
(\cite{wood99}).  Large-scale surveys also allow us to find very rare
objects, such as Cepheids in eclipsing binary systems
(\cite{alcock02}).

Thus motivated, we started a survey of the Local Group galaxy
Messier~33.  It is the ideal target since it has a large metallicity
gradient, is sufficiently nearby that we can obtain good-quality
photometry of most Cepheids, and it has a large number of Cepheids.

Messier~33 (M~33) can be considered as one of the stepping stones of
modern cosmology.  It is one of the first galaxies for which Hubble
obtained a distance, allowing him to show that many ``nebulae'' were
in fact galaxies similar to our own stellar system (Hubble, 1926).  A
number of surveys over the next few decades found variable stars in
M~33 (\cite{hs53,van75}). The last major photographic survey was by
\cite{kmw87} who discovered 90 Cepheids, over 50 Long Period
Variables, and several hundred unclassified variables.
More recently the DIRECT project conducted the first CCD variability
survey of M33, discovering more than a thousand variables (Macri et
al.~2001; Mochejska et al.~2001a,~2001b) in the central regions of
this galaxy.  \cite{shporer06} have also surveyed bright stars in M~33
over several seasons.
There is also a recent multi-color CCD-based survey by \cite{massey06}
covering a wide field.
These surveys all had one or several drawbacks, however, in particular
poor sensitivity (photographic), small field of view (recent CCD
surveys), or too few epochs.  The advent of wide-field cameras
(optical and near-infrared) on 4m-class telescopes makes such a survey
finally possible, without the drawbacks of previous observations.

\section{The CFHT survey}

We obtained data using MegaCam installed on the Canada-France-Hawai'i
Telescope.  It is made of 36 $2048\times 4612$ CCDs, offering a
one-square-degree field of view. Such a wide field covers the whole
galaxy, making it ideal for this survey. We used the $g^{\prime}$,
$r^{\prime}$ and $i^{\prime}$ Sloan filters, between August 2003 and
January 2005.  Between 33 and 36 images have been obtained in each
filter; we also have deep single-epoch $u*$ and $z^{\prime}$ data.
Exposure times were between 8 and 11 minutes depending on the filter.

\begin{figure}
\includegraphics[width=6cm]{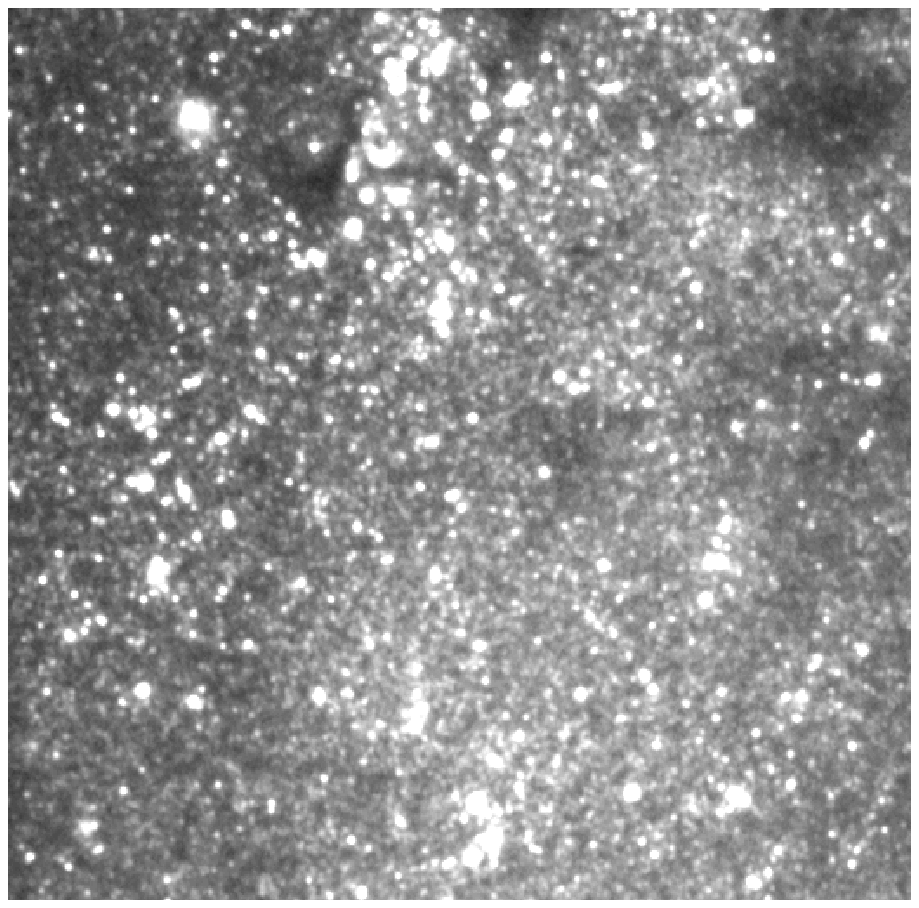}
\includegraphics[width=6cm]{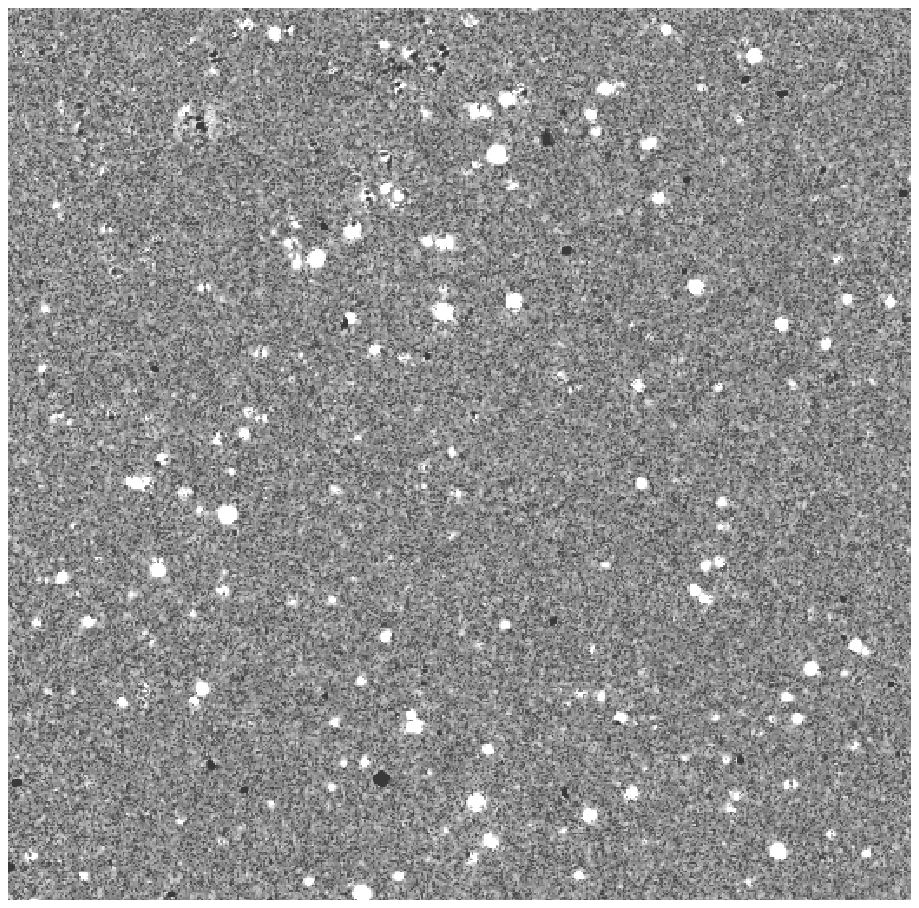}
\caption{{\it (Left)} Original image showing the level of crowding
near the center of M~33. {\it (Right)} After subtraction of the
reference image, only a few dozen objects are left. Note that variable
objects can appear as positive (white) or negative (black)
fluctuations, depending on whether the objects got brighter or fainter
between the two epochs.}
\label{fig:isis}
\end{figure}

The search for variables has been described in detail in
\cite{hartman06}. We will briefly recapitulate the procedure here.
The data reduction is done in two steps. First, the search for
variable objects is done using the image subtraction technique, as
implemented in the ISIS software (\cite{alard98}; \cite{alard00}).
All images are registered to a \emph{reference} image, obtained
by stacking several good-seeing images.  That reference image is
subtracted from each program image, after matching the point-spread
function and background. What is left is an image where most objects
have disappeared, except those that have varied (see
Fig.~\ref{fig:isis}).

For each variable object, image subtraction gives the number of counts
above or below the flux on the reference image. In other words it is
only a \emph{relative} flux. In order to calibrate the data on an
\emph{absolute} flux scale, we need to do point-spread function (PSF)
photometry of all sources. This is the second step of the processing.
This has been done using the DAOPHOT/ALLSTAR suite of programs
(Stetson 1987, 1992). The photometry was done on the reference image
in each filter.  Variable objects found with ISIS were matched to
point sources detected by DAOPHOT so that we can eventually obtain a
magnitude for each variable source.

\begin{figure}
\includegraphics[width=10cm]{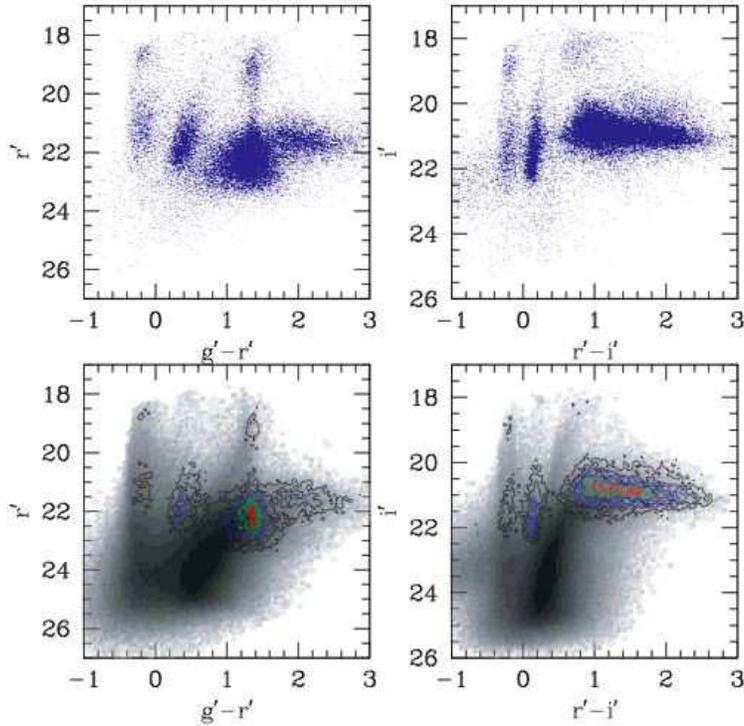}
\caption{{\it (Bottom)} Color-magnitude diagrams based on photometry
of the deep reference frames (grey scale). The contours indicate the
density of variables.  {\it (Top)} CMDs with variable stars only. They
are concentrated in several areas, most noticeably the main sequence,
the Cepheid instability strip, the tip of the red giant branch, the
asymptotic giant branch, and the red supergiant region.}
\label{fig:cmd}
\end{figure}

At the present time we do not have an accurate photometric
calibration. We can, however, use the default calibration provided by
the observatory. This gives us a reasonably good idea of stars'
magnitudes and colors. Knowing that, we can plot color-magnitude
diagrams (CMD), emphasizing the positions of variable objects (see
Fig.~\ref{fig:cmd}).

\section{Variable stars}

In total we have more than 36000 variable objects.  We performed rough
searches for known categories of variables. Several regions have been
cut out of the HR diagram, where we expect to find variable stars.
These are blue supergiant variables, red supergiant variables, Long
Period Variables (LPVs), and Cepheids.  The number of objects in each
group are given in Table~\ref{tab:stats}.

\begin{table} {\bf Statistics of variable stars in M~33} \\[1ex]
\begin{tabular}{l r}
\hline
Blue supergiants & 392   \\
Red supergiants  & 984   \\
LPVs             & 19767 \\
Cepheids         & 2327  \\
\hline
\end{tabular}
\label{tab:stats}
\end{table}

Note that there are thousands of other variable objects that are not
in any of these categories. As expected from inspection of
Fig.~\ref{fig:cmd}, LPVs represent the majority of all variable stars.
The fraction of variable objects on the asymptotic giant branch is
very large; actually in some limited magnitude range the majority of
AGB stars are variable (see Hartman et al.\ 2006 for details, in
particular their figure~6).

\subsection{Cepheids}

The number of stars in the region of the Cepheid instability strip is
3580.  However, the region defined as the Cepheid instability strip in
\cite{hartman06} is broad enough that it will necessarily contain
non-variable stars. It will also contain luminous main sequence objects
affected by strong reddening.  A substantial fraction do not show any
periodicity in the data, ruling them out as Cepheids.  This is why the
number of genuine Cepheids is smaller than 3580.  The obvious
discriminant between Cepheids and other types of stars is the period.
A search for periodicity has been done using the analysis of variance
method (Schwarzenberg-Czerny 1996).  In total, 2011 stars have a
robust period and can be considered as genuine Cepheids.  There is a
number of objects (a few hundreds) that are very good candidate
Cepheids but for which the period search didn't return a single, clear
period.  The total number of Cepheids is going to slightly increase as
we refine the analysis.

We then fitted Fourier series to the light curves.  The Fourier
parameters (see Simon \& Lee, 1981, for definition) are not very
precise owing to the limited number of epochs.  Nevertheless, one can
clearly recognize the pulsation mode for the majority of Cepheids.
According to the Fourier parameters (and to the position on the PL
diagram, see below), there are 1580 fundamental-mode pulsators, and
431 overtones.

\subsection{The metallicity of M~33}

M~33 was initially chosen because of its large metallicity gradient.
Recent measurements of the oxygen abundance [O/H] in H{\sc ii} regions
yield a conflicting result (e.g. Crockett et al. 2006).  Cepheids
themselves offer ways to constrain the metallicity of the young
stellar populations. The most effective way is through double mode
Cepheids.  The period ratio of beat Cepheids is very sensitive to the
metal content, as is known from analysis of beat Cepheids in the
Magellanic Clouds (e.g. Alcock et al. 1995).  Even with only $\sim 35$
epochs, it proved possible to find five double-mode Cepheids.  A
pulsation analysis of these five beat Cepheids shows that there is a
strong metallicity gradient (Beaulieu et al. 2006). A large gradient
is the only way to reconcile predictions of stellar pulsation theory
with the physical locations of beat Cepheids in M~33.

\begin{figure}
\includegraphics[width=10cm]{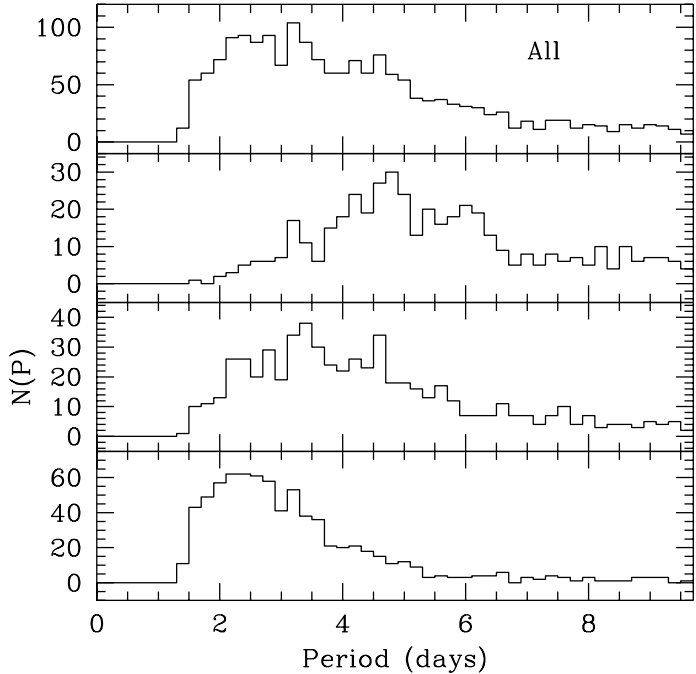}
\caption{Period distributions in three radial zones.  The peak of the
distribution is shifted to shorter and shorter periods as one gets to
larger and larger radii (from top to bottom). This is a clear effect
of a large metallicity gradient.}
\label{fig:pdist}
\end{figure}

Metallicity also affects the Cepheid period distribution, in the sense
that as [O/H] goes down, the peak of the Cepheid period distribution
moves to shorter periods. A plot of the Cepheid period histogram
(Fig~\ref{fig:pdist}) clearly shows this.

\subsection{Period--Luminosity relation}

\begin{figure}
\includegraphics[width=10cm]{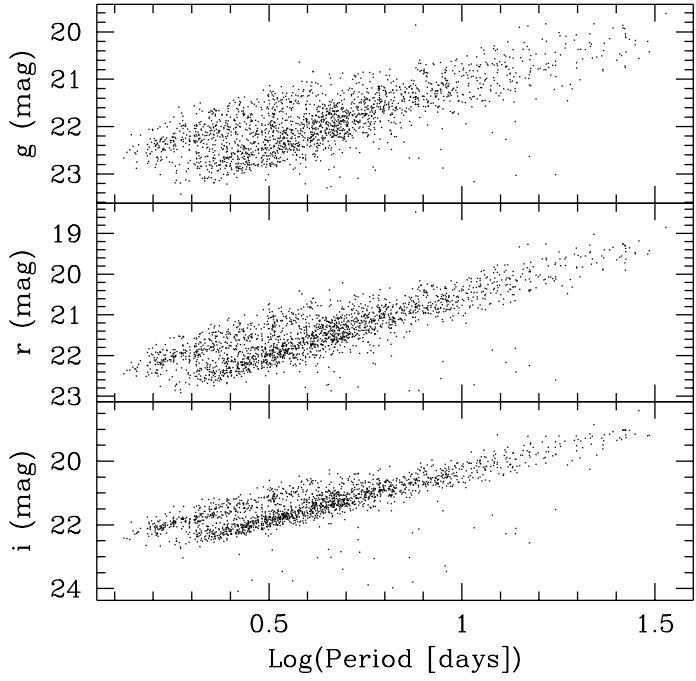}
\caption{Period-Luminosity relations in $g^{\prime} r^{\prime}
i^{\prime}$ filters. One can clearly distinguish the fundamental and
first overtone sequences. We can also see a few Population II
variables.}
\label{fig:pl}
\end{figure}

We need to emphasize the fact that the photometric calibration is to
be taken as temporary. We have noticed significant differences in
photometric zero point from CCD to CCD and it would be dangerous to
use these data if a robust zero point is needed (in particular for
distance scale work).  We are currently in the process of acquiring
calibration data in the Sloan filters and in $UBVRI$. This will allow
us to calibrate the CFHT data and also to produce Sloan-to-$UBVRI$
transformations adequate for Cepheids.
This being said, the calibration is already good enough for the
Period--Luminosity relation to show clearly the two sequences expected
for fundamental-mode and first overtone pulsators respectively (see
Fig~\ref{fig:pl}).

\section{Beyond a survey}

While this survey is motivated by the effect of metallicity on the
Cepheids PL relation, most variable objects are not Cepheids:  the
majority of variables are red giants. While we do not have at present
enough data to find periods for LPVs, we plan to add more data so we
can determine periods for these stars. This would greatly enhance the
legacy value of the survey.

These data can also be useful for very rare objects. The candidate
black hole binary M~33 X-7 is visible in our data, and its light curve
has already been analyzed by \cite{shporer07}.

M~33 has also been surveyed at other wavelengths (Spitzer, Chandra,
XMM) and results have appeared recently (e.g. McQuinn et al 2007;
Grimm et al. 2005). Cross-correlation of the various catalogs of
objects at different wavelengths, coupled with variability
information, opens a new way of studying stellar populations in nearby
galaxies.

Finally, we note that all the information contained in the tables in
\cite{hartman06} is publicly
available\footnote{http://www.astro.livjm.ac.uk/$\sim$dfb/M33/}.  This
web site is kept up to date regarding the calibration and the progress
of the analysis.

\end{document}